\documentclass[journal,twoside,letterpaper]{IEEEtran}
\newcommand{\numcol}{2}

\usepackage{setspace}

\usepackage{verbatim}

\usepackage{amsmath}
\usepackage{amssymb}
\usepackage{graphicx}
\graphicspath{{eps/}}
\usepackage[noadjust]{cite}
\usepackage{subfigure}

\usepackage{ifthen}
\newcommand{\checkOneCol}{\ifthenelse{\numcol = 1}}

\newcommand{\graphwidth}{\columnwidth}%
\usepackage{url}
\begin{document}

\newcommand{\papertitle}{Self-Organization of Wireless Ad Hoc Networks as Small Worlds Using Long Range Directional Beams}


\title{\papertitle}

\author{\authorblockN{Abhik Banerjee\authorrefmark{1}, Rachit Agarwal\authorrefmark{2}, Vincent Gauthier\authorrefmark{2}, Chai Kiat Yeo\authorrefmark{1}, Hossam Afifi\authorrefmark{2} and Bu Sung Lee\authorrefmark{1} \\}
\authorblockA{\authorrefmark{1} CeMNet, School of Computer Engineering, Nanyang Technological University, Singapore\\}
\authorblockA{Email: \{abhi0018, asckyeo, ebslee\}@ntu.edu.sg\\}
\authorblockA{\authorrefmark{2}Lab. CNRS SAMOVAR UMR 5157, Telecom Sud Paris, Evry, France\\}
\authorblockA{Email: {\{rachit.agarwal, vincent.gauthier, hossam.afifi\}}@it-sudparis.eu}
}

\maketitle

\begin{abstract}
We study how long range directional beams can be used for self-organization of a wireless network to exhibit small world properties. Using simulation results for randomized beamforming as a guideline, we identify crucial design issues for algorithm design. Subsequently, we propose an algorithm for deterministic creation of small worlds. We define a new centrality measure that estimates the structural importance of nodes based on traffic flow in the network, which is used to identify the optimum nodes for beamforming. This results in significant reduction in path length while maintaining connectivity.
\end{abstract}
\section{Introduction}\label{sec:Intro}
Initially studied in the context of social networks \cite{MilgramSW}, the small world phenomenon refers to the ability of such networks to maintain high clustering and low average path length between nodes. Subsequent research has explored ways to redesign networks so as to exhibit small world characteristics. Watts \& Strogatz \cite{WattsStrogSWN} showed that regular networks can be reorganized to exhibit small world behaviour by randomly rewiring a small set of links.

Small worlds are an attractive model for reorganizing a wireless ad hoc network so as to ensure performance guarantees. As wireless networks typically suffer from issues of connectivity, maintaining high clustering guarantees reliability. Further, reorganization of a network with the path length bounded as the logarithm of the network size ensures performance scalability. However, due to the spatial nature of wireless networks, links between nodes cannot be randomly rewired as they are constrained by the transmission range. In \cite{HelmySWWi}, Helmy used simulation results to study the behaviour of wireless networks as a result of random addition of distance limited short cuts. Instead of random rewiring or addition of links, deterministic placement of short cuts was studied by \cite{ChitraWired} and \cite{SharmaHSN} in which they consider hybrid sensor networks that include a small set of wired links. Similarly, Guidoni et al. in \cite{GuidoniHetSN} proposed using high capacity nodes in a heterogenous sensor network for short cut creation.

In this paper, we study how small world behavior can be realized in a wireless network by the use of directional beam forming at nodes. Our primary motivation for using directional antennas stems from the fact that they can be used to transmit over longer transmission ranges than omnidirectional antennas while using the same transmission power. This implies that short cuts can be created between nodes without the need for additional infrastructure. This distinguishes us from existing literature that focus on addition of new links between nodes. Instead, existing omnidirectional links are rewired as long range directional ones. Another unique property of using directional antennas is that the shape of the beam implies that long range links can be created with more than just one node. Thus, the fraction of long range links in the network is actually higher than the number of nodes beamforming. The focus of this paper is to study the issues surrounding the use of directional beamforming for small world creation in a wireless ad hoc network and propose ways to achieve an optimal design. To our knowledge, the only other work where directional antennas were used for short cut creation was in \cite{VermaSWWMN}. However, like other existing papers, the proposed model considers addition of links using multiple radios.

The first part of the paper provides a simulation based analysis of using randomized directional beamforming for small world creation in wireless ad hoc networks. Our results show that, while significant reduction can be achieved in the average path length, this is accompanied by loss in connectivity. Subsequently, we focus on distributed algorithm design for deterministic small world creation. Our design centers on a new definition of centrality that is computed individually at nodes based on information obtained from overhearing traffic flows in the network. We show that significant reduction in path length can be achieved while maintaining connectivity. Our focus here is that of a connected network with the primary motivation of limiting the growth in path length with increasing size of the network without suffering losses in connectivity. A companion paper in the same workshop \cite{AgarwalBioIns} considers a disconnected network and how bio-inspired mechanisms can be used to ensure connectivity by using directional beams.




\section{Small World Wireless Networks using Directional Antennas}\label{sec:swnda-anal}
In this section, we do a simulation based analysis of using directional antennas for small worlds in wireless networks. We use the results to identify crucial design aspects.

\subsection{Network Model}\label{subsec:nw-model} 
We consider a wireless ad hoc network of $N$ nodes all of which consist of a single beamforming antenna. Initially, all nodes transmit using omnidirectional beams with range $r$. Subsequently, a fraction $p$ of the nodes in the network are randomly chosen which use long range directional beams. Usage of directional antenna by a node can be classified into different categories depending on the modes of tranmission and reception \cite{LiAsymConn}. We consider that when a node creates a long range directional beam, it operates in the mode of directional transmission and omnidirectional reception (DTOR). For the purpose of analysis, we use the sector model for directional antennas \cite{YuConnCov}. We compare the simulation results using the sector model to a more realistic uniform linear array (ULA) antenna \cite{BettstetterRndBeamform} model.

\begin{figure}[tb]
    \centering
    \includegraphics[width=\graphwidth]{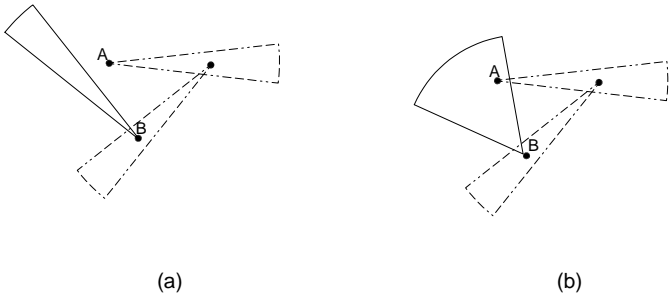}
    \caption{Effect of beam width on the connectivity. Using a wider beam with shorter beam length allows a bidirectional path between nodes $A$ and $B$.}
    \label{fig:conn_wid}
\end{figure}

A directional beam is characterized by the beam length, width and the beam direction. For the ULA, a longer beam length can be achieved by increasing the number of antenna elements used \cite{BettstetterRndBeamform}. By keeping the transmission power constant, increasing the number of elements results in a narrower and longer main lobe, while using a single antenna results in an omnidirectional beam. When using the sector model as an abstraction, a constant transmission power implies that the area covered by the beam stays constant. A beam of width $\theta$, therefore, results in a beam length $r(\theta) = r \sqrt{\frac{2 \pi}{\theta}}$. The directed nature of the beam leads can lead to the problem of asymmetric paths between nodes, as discussed in \cite{LiAsymConn,YuConnCov}. As illustrated in Fig. \ref{fig:conn_wid}, bidirectional connectivity between two nodes essentially requires the presence of a circular path.

To account for the above tradeoffs, we choose a beamwidth that optimizes connectivity and beamlength depending on the density of nodes. To incorporate the tradeoff between increased length and connectivity, we divide the sector into separate regions of width $r$. Subsequently, we weigh the beam length $r(\theta)$ with the probability that at least one node is located in the first and the last regions,
\begin{equation}\label{eq:lC}
r_C = r(\theta) p_{nf} p_{nl}
\end{equation}
where $p_{nf}$ and $p_{nl}$ are the probabilities that at least one node is located in the first and last sectoral regions respectively. The term $p_{nf}$ is indicative of the probability that the node maintains connectivity to the nodes in its omnidirectional neighbourhood while $p_{nl}$ indicates the probability that at least one node benefits from the increased beam length. The optimum beam width $\theta^{*}$ among a set of values for $\theta$ is chosen as the one that maximizes the weighted beam length $r_C$, i.e.
\begin{equation}\label{eq:lcondn}
\theta^{*} = \operatorname*{arg\,max}_{\theta} \left[r \sqrt{\frac{2 \pi}{\theta}}\right] p_{nf} p_{nl}
\end{equation}
The values for $p_{nf}$ and $p_{nl}$ are obtained based on the node density in the network. Given that the number of nodes in the omnidirectional neighbourhood of a node is $n$, the corresponding values are obtained as $p_{nf} = 1 - (1 - \frac{A_f}{\pi r^2})^n$ and $p_{nl} = 1 - (1 - \frac{A_l}{\pi r^2})^n$ where $A_f$ and $A_l$ are the area of the first and last regions respectively. Recall that the area under the beam is equal to that of the omnidirectional area when the same transmission power is used.

While the sector model is ideal for analysis, we also run simulations on a more realistic model of directional antennas to compare the performance. The model we use is that of a uniform linear array (ULA) \cite{BettstetterRndBeamform}, in which the antenna elements are arranged linearly. The beam pattern of a ULA is characterized by the number of elements used $m$ and the boresight direction $\theta_b$. An important difference with the sector model is that the beam pattern of ULA is characterized by one or more side lobes in addition to the main lobe. The maximum gain obtained in the direction of $\theta_b$ is equal to $m$. In order to map the sector model used earlier to ULA, we equate the number of antenna elements $m = r(\theta^{*})$ where $r(\theta^{*})$ is the normalized beam length corresponding to $\theta^{*}$ obtained in equation (\ref{eq:lcondn}).

\begin{figure}[tb]
    \centering
    \subfigure[Path Length and Clustering Coefficient]{
    \includegraphics[width=\graphwidth]{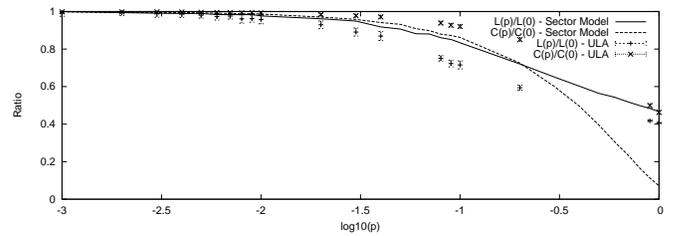}
    \label{subfig:aplcc_varpn300}
    }

    \subfigure[Unidirectional Connectivity]{
    \includegraphics[scale=0.8]{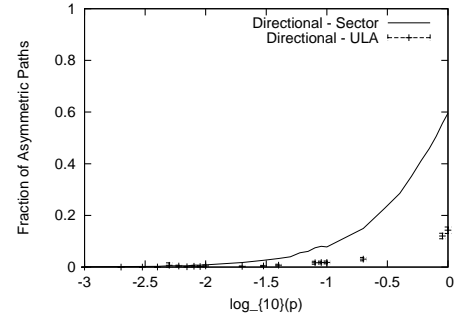}
    \label{subfig:uni_varpn300}
    }

    \caption{Small World characteristics as a function of varying probability of rewiring for $N=300$. $L(0)$ and $C(0)$ denote the average path length and clustering coefficient for the initial network with all nodes using omnidirectional antennas.}
    \label{fig:varp_n300}
\end{figure}

\subsection{Simulation Results}\label{subsec:sim-results}
For our simulations, we consider a network consisting of nodes using omnidirectional antennas distributed randomly in a rectangular region. We study the impact of using randomly oriented directional beams on the average path length (APL) and connectivity of this network. 

For the first set of simulations, we vary the fraction of nodes $p$ that use directional beams while the rest of the nodes continue to use omnidirectional beams. The omnidirectional transmission range is normalized to $1$ with nodes distributed randomly in a $10$x$10$ region. Fig. \ref{fig:varp_n300} shows the impact on path length reduction and connectivity for $N=300$. The results for the sector model, shown using lines, are compared to those using the ULA model, shown with dots. The beam length $r(\theta^{*})$ obtained using equation (\ref{eq:lcondn}) results in a ratio $\frac{r(\theta^{*})}{D} \approx 0.2$. $D$ denotes the maximum distance between any two nodes in the network, i.e. the diameter of the network since we normalize $r$ to $1$. In Fig. \ref{subfig:aplcc_varpn300}, the ratio of the reduced path length $L(p)$ and clustering coefficient $C(p)$ to the initial values $L(0)$ and $C(0)$ are shown. We note that the values for $\frac{L(p)}{L(0)}$ and $\frac{C(p)}{C(0)}$ are quite close to each other for low values of $p$ when using the sector model. This is contrary to the desired results for small world networks as the path length reduction is accompanied by loss in connectivity. The adverse impact of $C(p)$ on connectivity is seen in Fig. \ref{subfig:uni_varpn300} as a high percentage of nodes have asymmetric paths. However, when a more realistic ULA model is used, better results are obtained in terms of both the path length improvement and connectivity. The presence of side lobes implies that a beamforming node retains connectivity to a greater fraction of nodes in its omnidirectional neighbourhood, resulting in higher values of the clustering coefficient. This also accounts for shorter path lengths since, in contrast to the sector model, these nodes in the omnidirectional neighbourhood can now be reached in a single hop.

\begin{figure}[tb]
    \centering
    \includegraphics[scale=0.8]{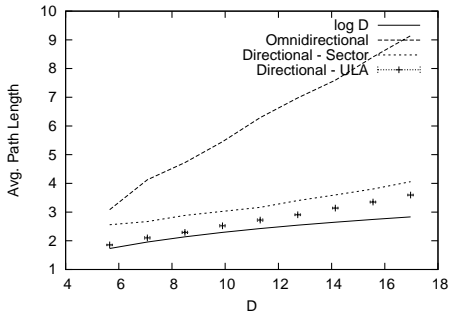}
    \caption{Growth of APL with increase in the size of the simulation region.}
    \label{fig:apl_vardiam}
\end{figure}

To underscore the suitability of directional beamforming for small world creation, we view it in the context of existing literature. Helmy \cite{HelmySWWi} obtained results showing the impact of distance limited short cuts on the path length reduction. In our results when using the sector model, for the corresponding value of $\frac{r(\theta^{*})}{D}$, we observe that path length reduction is comparable for small values of $p$ though the reduction is less for higher values. For the more realistic ULA model, though, the path length improvements can actually exceed those of \cite{HelmySWWi}. We also run additional simulations to study the growth in path length compared to the logarithm of the network size, which is measured in terms of $D$ as this is a spatial network. Fig. \ref{fig:apl_vardiam} shows that the APL grows in $O(\log D)$ when $p=1$.


\subsection{Discussion and Insights}\label{subsec:insight}
An important feature of our results in the previous section is the tradeoff between path length improvement and connectivity in the network. Despite choosing beam width so as to maintain connectivity, we observe that a high fraction of beamforming nodes drastically reduces the probability of having a bidirectional path between nodes. Based on the results from sector model alone, it is difficult to identify an optimum value of $p$ such that the network exhibits small world behaviour. However, looking at the results using the realistic ULA model, we observe that the constraints imposed by the sector model can be relaxed to an extent. We observe in Fig. \ref{fig:varp_n300} that a $30 \%$ of reduction in the average path length can be achieved with $p = 0.1$ while less than $2 \%$ of node pairs suffer from unidirectional connectivity. This also provides us with additional insight for algorithm design. While the sector model is more tractable for algorithm design, we conclude that an algorithm that results in $20 \%$ of links being unidirectional is permissible as that is compensated in a realistic setup.

\section{Traffic Aware Small World Creation using Directional Beamforming}\label{sec:tfaware-sw}
We outline an algorithm design for realizing small world behaviour by deterministically choosing the set of nodes to beamform and the beamforming parameters. Given our primary objective of maximizing the reduction in path length across the network, we need to identify the set of nodes that are most ideally located to minimize path lengths across the network. This correlates directly to the traditional notion of betweenness centrality \cite{FreemanBw}. Creating short cuts on nodes with high values of betweenness is likely to reduce path lengths across the network as they lie along the majority of shortest paths between nodes in the network. For distributed self-organization of the network as a small world network, a measure of betweenness needs to be identified that can not only be computed in a distributed manner but one that adapts to the routes taken by flows. Further, as nodes need to decide on their beamforming behaviour individually, they also need to be able to estimate their rank of their betweenness in the context of the network.

We propose the Wireless Betweenness Centrality (WFB) which is computed at nodes based on neighbourhood traffic flow information. A key aspect of WFB is that it exploits the Wireless Broadcast Advantage (WBA) as nodes use implicitly available information from neighbourhood transmissions to compute their centrality values. WFB values are computed recursively allowing network structural information to propagate over multiple hops. Subsequently, we propose an algorithm that uses WFB values at nodes along with insights obtained from simulation results to realize small world behaviour in the network.

\subsection{Wireless Flow Betweenness (WFB)}\label{subsec:wfb-defn}
The broadcast nature of the wireless medium results in implicit sharing of information among nodes in the network. This feature is known as the wireless broadcast advantage (WBA) \cite{WieselthierWMA} and has been exploited in existing literature for improving network performance. We note that, as each node is aware of all transmissions in its neighbourhood, it can estimate the set of source-destination pairs corresponding to the observed flows. As a result, a simple measure of betweenness of a node may be obtained as,
\begin{equation}\label{eq:bw-simp}
Bet(v) = \frac{g(v)}{\sum\limits_{u \in \{\mathcal{N}(v) \cup v\}} g(u)}
\end{equation}
where $g(u)$ denotes the number of packets forwarded by a node $u$ for distinct source-destination pairs while $\mathcal{N}(v)$ denotes the set of neighbours of $v$. The denominator of the above expression gives the total number of packets forwarded in the neighbourhood of $v$. Each node records the number of packets forwarded by each of its neighbours.

However, the value of $Bet(v)$ computed using (\ref{eq:bw-simp}) does not give any idea about the importance of node $v$ with regard to the structure of the rest of the network. In order to do so, we propose exploiting WBA to propagate information over multiple hops. Our design proceeds by having nodes broadcast their self computed values of centrality as part of packet transmissions. Whenever a node transmits a packet, either as part of forwarding it for another flow or as a source, it appends the value of centrality computed for itself to the packet. Any neighbour overhearing this transmission stores the centrality value in addition to updating the number of packets forwarded by this node.

We elaborate on the recursive computation of Wireless Flow Betweenness (WFB) for a node $v$ and its neighbours when the former transmits a packet in a time slot $t$. Let the WFB of a node $v$ after $(t-1)$ time slots be denoted as $w^{t-1}(v)$. For each node $u \in \mathcal{N}(v)$, $v$ stores the number of packets forwarded along with their WFB values. When transmitting a packet in the $t$th time slot, $v$ updates its WFB using the following expression and appends it to the transmitted packet,
\begin{equation}\label{eq:wfb-fwd}
w^t(v) = \frac{g^t(v)}{w^{t-1}(v) \sum\limits_{u \in \{\mathcal{N}(v) \cup v\}} \frac{g^{t-1}(u)}{w^{t-1}(u)}}
\end{equation}
Upon receiving the value of WFB transmitted by $v$, each of its neighbours $u \in \mathcal{N}(v)$ recomputes its own value of WFB as,
\begin{equation}\label{eq:wfb-nbr}
w^t(u) = \frac{g^t(u)}{w^{t-1}(u) \big[\frac{g^t(v)}{w^t(v)} + \sum\limits_{u' \in \{\{\mathcal{N}(u) \backslash v\} \cup u\}} \frac{g^{t-1}(u')}{w^{t-1}(u')}\big]}
\end{equation}
Note that, for all nodes apart from $v$, the WFB recorded till the $(t-1)$th time slot is used. This is because, transmission from any node $u' \in \{\mathcal{N}(u) \backslash v\}$ in the time slot $t$ would imply a collision at $u$. Since we assume successful reception of the transmission of $v$ at $u$, it precludes the above condition.

It is easy to see how this recursive computation results in propagation of information regarding centrality across the network. After $v$ transmits in the $t$th time slot, its neighbours $u \in \mathcal{N}(v)$ are scheduled in subsequent time slots in which they broadcast their updates values of WFB leading to further recomputation.

\begin{figure}[tb]
    \centering
    \subfigure[Initial Omnidirectional Network]{
    \includegraphics[scale=0.4]{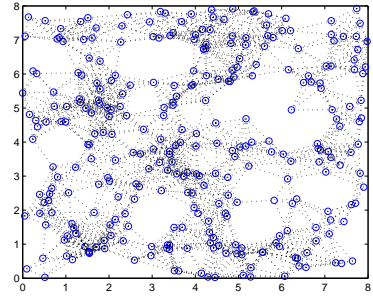}
    \label{subfig:nw_omni}
    }

    \centering
    \subfigure[After nodes use self computed values of WFB to decide on directional antenna behaviour. The solid edges correspond to directional beams.]{
    \includegraphics[scale=0.4]{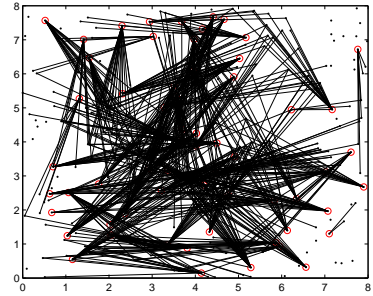}
    \label{subfig:nw_dir}
    }

    \caption{Using WFB to achieve small properties in a wireless network when $50 \%$ of all nodes transmit a packet to randomly chosen destinations.}
    \label{fig:nw_ill_wfb}
\end{figure}

\subsection{Directional Beamforming using WFB}\label{subsec:beamf-wfb}
Our next challenge is to have nodes decide on using long range directional beams based on their individually computed values of WFB. From the point of view of centrality, we observe that a node with a high value of centrality is likely to have neighbours with high values as well. This is particularly true in our scenario since the WFB value of a node is weighted with the values of its neighbours. However, this is also true for a generic view of a network due to the fact as nodes with high values of closeness centrality are likely to be clustered towards the center of the network. In addition to identifying the nodes with high values of WFB, we ensure that the connectivity of beamforming nodes is maintained by spacing out such nodes. This ensures that a beamforming node has a high percentage of omnidirectional nodes in its neighbourhood which preserve bidirectional connectivity.

Based on the above observations, we propose an design in which a node $v$ decides on using a directional beam if the average WFB of its neighbourhood is close to its own value of WFB, as determined by a similarity constant $\beta$. Thus, a node $v$ decides to use a directional beam after $t$ time slots if
\begin{equation}\label{eq:beam-condn}
|w^t_{avg}(\mathcal{N}(v)) - w^t(v)| < \beta w^t(v)
\end{equation}
where $w^t_{avg}(\mathcal{N}(v))$ is the average WFB of all nodes $u \in \mathcal{N}(v)$. Once a node decides to use a directional beam, it broadcasts its decision over both its original omnidirectional neighbourhood as well as the new neighbourhood determined by the directional beam. Subsequently, any node overhearing either of these two broadcasts decides against using a long range beam even if the above condition is satisfied for the node itself. This ensures that nodes using long range links are separated sufficiently so as to maintain connectivity.

Having decided on using a directional beam, a node needs to identify the optimal beam width and direction. The choice of beam width $\theta$ is done in the same manner as earlier, based on the node density using equation (\ref{eq:lcondn}). Instead of using a random direction, though, a node orients its beam in a direction in which it records the maximum hop count. This ensures that the longest paths in the network benefit from the self-organization. Fig. \ref{fig:nw_ill_wfb} illustrates the creation of long range directional beams in a network based on the WFB values computed at nodes when $50 \%$ of all nodes transmit a packet to randomly chosen destinations.

\begin{figure}[tb]
    \centering
    \subfigure[Reduction in Avg. Path Length]{
    \includegraphics[scale=0.8]{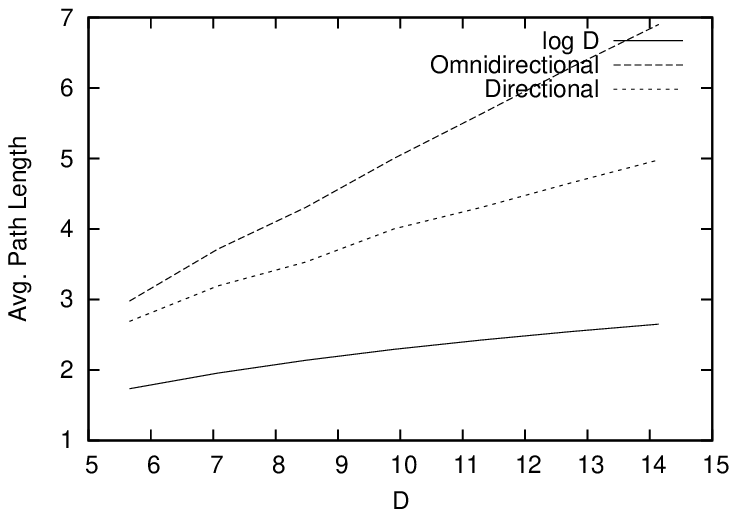}
    \label{subfig:fpl_f05}
    }

    \centering
    \subfigure[Unidirectional connectivity]{
    \includegraphics[scale=0.8]{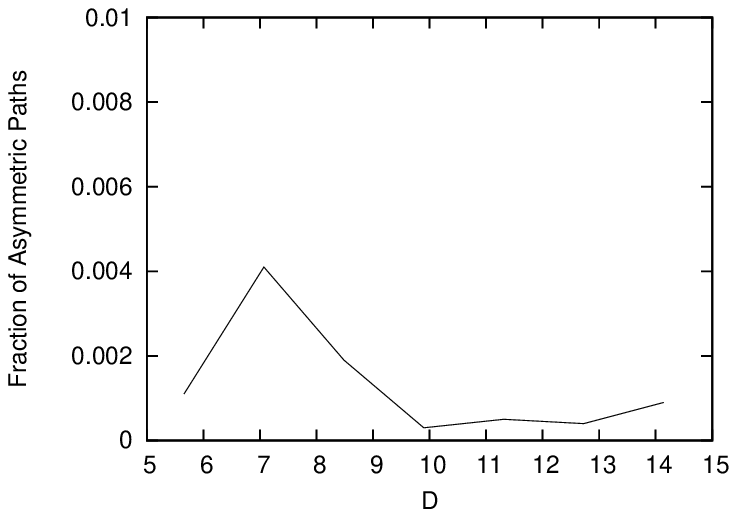}
    \label{subfig:udconn_f05}
    }

    \caption{Path Length and Connectivity for self-organization using WFB.}
    \label{fig:vardiamwfb_f05}
\end{figure}

\subsection{Simulation Results}\label{subsec:sim-wfb}
We evaluate the proposed algorithm using MATLAB simulations. We consider a dense network in which connectivity is guaranteed for the initial network setup in which all nodes use omnidirectional beams. The current set of results only consider the sector model of directional antennas. Nodes are randomly distributed over a square region. We obtain results for increase size of the network area, or the diameter $D$. Fig. \ref{fig:vardiamwfb_f05} shows the results when $50 \%$ of nodes generate source packets to randomly chosen destination nodes. We obtain a beam length ratio $\frac{r(\theta^{*})}{D} \approx 0.3$ while the fraction of nodes $p$ lies between $0.04$ and $0.07$. A path length reduction of up to $30 \%$ is achieved as the network size increases. Further, the effect on connectivity is negligible as less than $0.4 \%$ of all node pairs have asymmetric paths. The path length improvements are equivalent to those obtained by Helmy \cite{HelmySWWi} for corresponding values of $\frac{r(\theta^{*})}{D}$. Further, the path length improvements are much higher than corresponding values of $p$ for randomized beamforming shown earlier. We observe that the reduction in path length is lower than that shown in Fig. \ref{fig:apl_vardiam} even though the growth in the average path length is in $O(\log D)$. The difference in performance is explained by the fact that the results shown in Fig. \ref{fig:apl_vardiam} correspond to $p=1$ which is much higher than what is obtained here. The results shown are for $\beta = 0.2$. Using higher values of $\beta$ results in greater reduction in path length but is accompanied by loss in connectivity as a greater number of neighboring nodes use directional beams. However, higher values of $\beta$ can be used for realistic beamforming models such as ULA as they result in greater connectivity. We leave this for future investigation.

\section{Conclusion}\label{sec:concl}
In this paper, we explore the use of directional beamforming for self-organization of a dense wireless ad hoc network as a small world. We provide a simulation based analysis of relevant issues in such a scenario followed by an algorithm design for deterministic small world creation. Our results show that significant path length reduction can be achieved with negligible effect on the connectivity. As part of our future work, we would like to extend the algorithm design to achieve greater reduction in path length.

\end{document}